# Machine learning enables polymer cloud-point engineering *via* inverse design

Jatin N. Kumar,[1,*] Qianxiao Li,[2] Karen Y.T. Tang,[1] Tonio Buonassisi,[3] Anibal L. Gonzalez-Oyarce,[2] Jun Ye,[2]

[1]Institute of Materials Research & Engineering, 2 Fusionopolis Way, #08-03, Singapore 138634

[2]Institute of High-Performance Computing, 1 Fusionopolis Way, #16-16, Singapore 138632

[3]Massachussets Institute of Technology, Cambridge, MA 02139, USA

**Inverse design is an outstanding challenge in disordered systems with multiple length scales such as polymers, particularly when designing polymers with desired phase behavior. We demonstrate high-accuracy tuning of poly(2-oxazoline) cloud point *via* machine learning. With a design space of four repeating units and a range of molecular masses, we achieve an accuracy of 4°C root mean squared error (RMSE) in a temperature range of 24–90°C, employing gradient boosting with decision trees. The RMSE is >3x better than linear and polynomial regression. We perform inverse design *via* particle-swarm optimization, predicting and synthesizing 17 polymers with constrained design at 4 target cloud points from 37 to 80°C. Our approach challenges the status quo in polymer design with a machine learning algorithm, that is capable of fast and systematic discovery of new polymers.**

Polymers are ubiquitous in both structural and functional systems, owing to their highly tunable physical, chemical, and electrical properties.[1-4] The development of polymers has historically been based on an Edisonian approach. Herein, we develop a machine learning framework to predict polymer structure (topology, composition, functionality, and size), on the basis of target phase properties, specifically the cloud point. This framework accommodates the complex disorder across multiple length scales that distinguishes polymers from small molecules, [5-7] inorganic crystals,[8] and systems-structure optimization.[9-11]

Phase properties, which describe the order of a polymer across multiple length scales, are determined by interactions of polymers with the solution and themselves. One such phase property is the cloud point, the temperature at which polymers are no longer miscible in solution. Numerous studies tabulate simple relationships between cloud point and one or two experimental variables (*e.g.*, structure[12] and temperature[13,14]), or offer polynomial fits to the data.[15] Ramprasad and colleagues applied machine learning to density-functional theory (DFT) calculations to predict opto-electronic[16,17] and physical[18] bulk polymer properties.[4,18] However, this approach is computationally expensive,[19,20] particularly for polymer systems,[21]

and does not enable scalable inverse design over a wide range of conditions with high accuracy.[22,23]

In this study, we combine machine learning, domain expertise, and experiment to solve the inverse design problem for polymers. Our framework (**Figure 1**) has 3 parts: (1) data curation (defining material descriptors) that relates poly(2-oxazolines) cloud point, size, and relative ratios of 4 different monomer units; (2) machine learning algorithm selection and hyperparameter tuning to enable fast forward prediction of cloud point based on structure with the evaluation of algorithmic robustness over systematic error and differing data quality[24]; and (3) use of said algorithm for inverse design using particle swarm optimization (PSO) with design selection using an ensemble of neural networks. We demonstrate the accuracy of our inverse-design paradigm by predicting the compositions of, and synthesizing, 17 polymers with cloud points between 37 to 80°C, using a modular combination of 4 repeating monomer units. We achieve ~ 4°C error, nearly within experimental error (1–3°C).

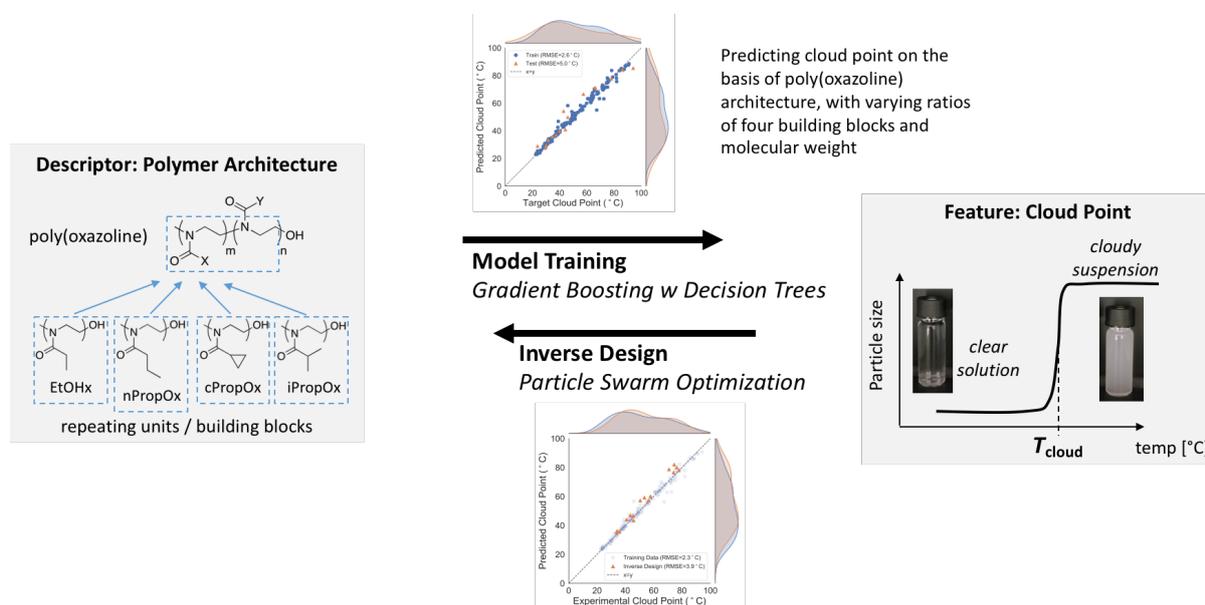

**Figure 1: Study framework.** First, we train a machine learning model to predict cloud point on the basis of poly(2-oxazoline) structure, with varying ratios of four monomer units (building blocks) and molecular weights. Second, we demonstrate inverse design using the trained algorithm and particle swarm optimization, predicting 17 polymer structures from user-defined cloud points. The model accommodates the inherent complexity of polymers over multiple length scales.

## Results & Discussion

We combine and curate literature and experimental data to create the input into our machine learning framework. Historical cloud-point data for poly(2-oxazoline)s[15, 25-30] was curated into a set of input variables ((1) molecular weight of the polymers; (2) polydispersity index; (3) polymer type (homo, statistical, or block); (4) total number of each monomer unit in the final polymer (A: EtOx, B: nPropOx, C: cPropOx, D: iPropOx, E: esterOx)) and output variables (cloud point in ˚C) (**Table S1**). We synthesized a series of poly(2-oxazoline)s by similar methods to augment this data (**Table S2**). Cloud point was evaluated by dynamic-light scattering (DLS) in accordance with best practices,[31] particularly since DLS affords greater weightage to the modal mass as a correction for the unsymmetric molecular weight distributions (MWD) of our synthesized polymers (details in ESI). Due to data scarcity, esterOx was not synthesized nor considered in inverse design. While a general relationship of input variables to output could be observed from **Figure 2**, it is well documented that machine learning methods generally have superior predictive accuracies in multi-variable parameter space. [32-34]

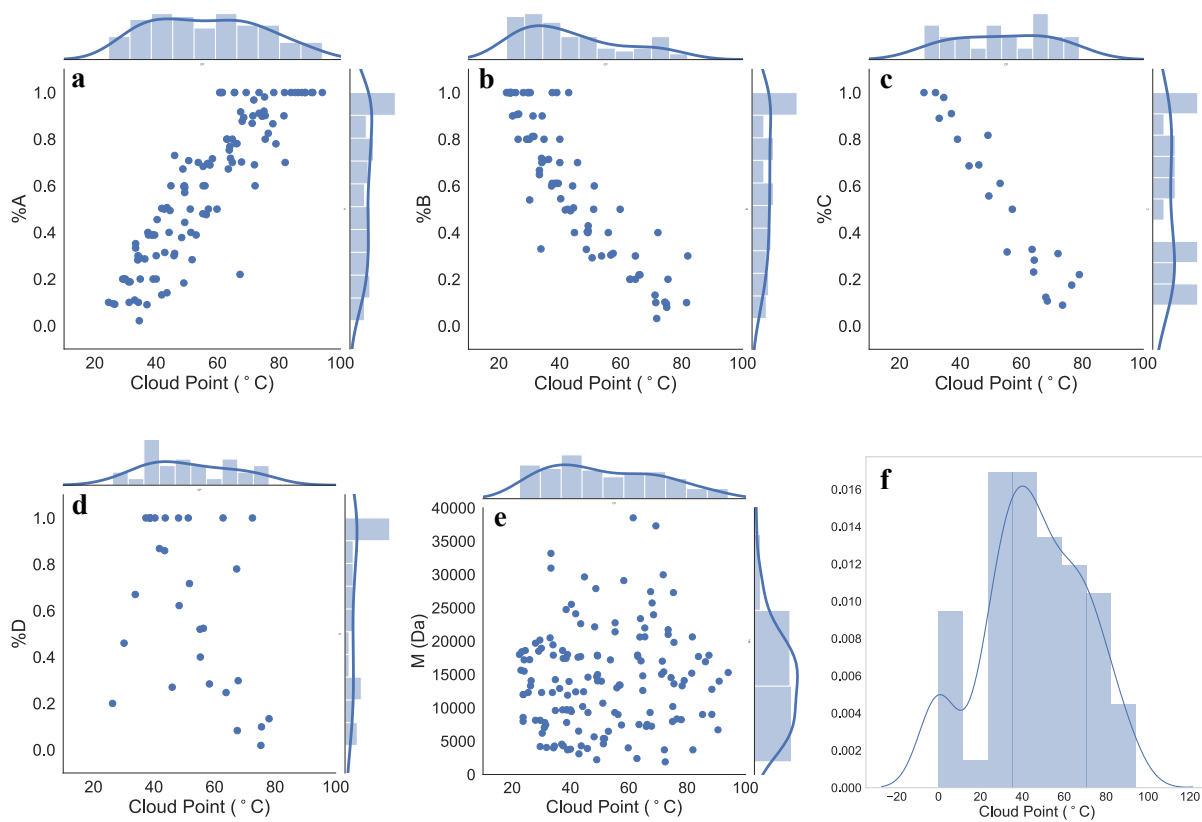

**Figure 2:** The dependencies cloud points to the mole fraction of (a) EtOx; (b) nPropOx; (c) cPropOx; (d) iPropOx; and (e) molecular mass (M), where all zero values were filtered from the graphs, and, (f) the number distribution of cloud-point where zero represents polymers without a CP

We compared the root-mean-squared errors (RMSE) of simple linear and quadratic regressions against machine learning methods including support vector regressions (SVR), neural networks (NN) and gradient boosting regression with decision trees (GBR) (**Figures 3, S3**). The accuracies of the various models were determined by splitting the input dataset into training, validation, and test sets, with training and validation performed from historical data, while testing was performed with experimental data. The RMSE and inference times are reported in **Table S3**.

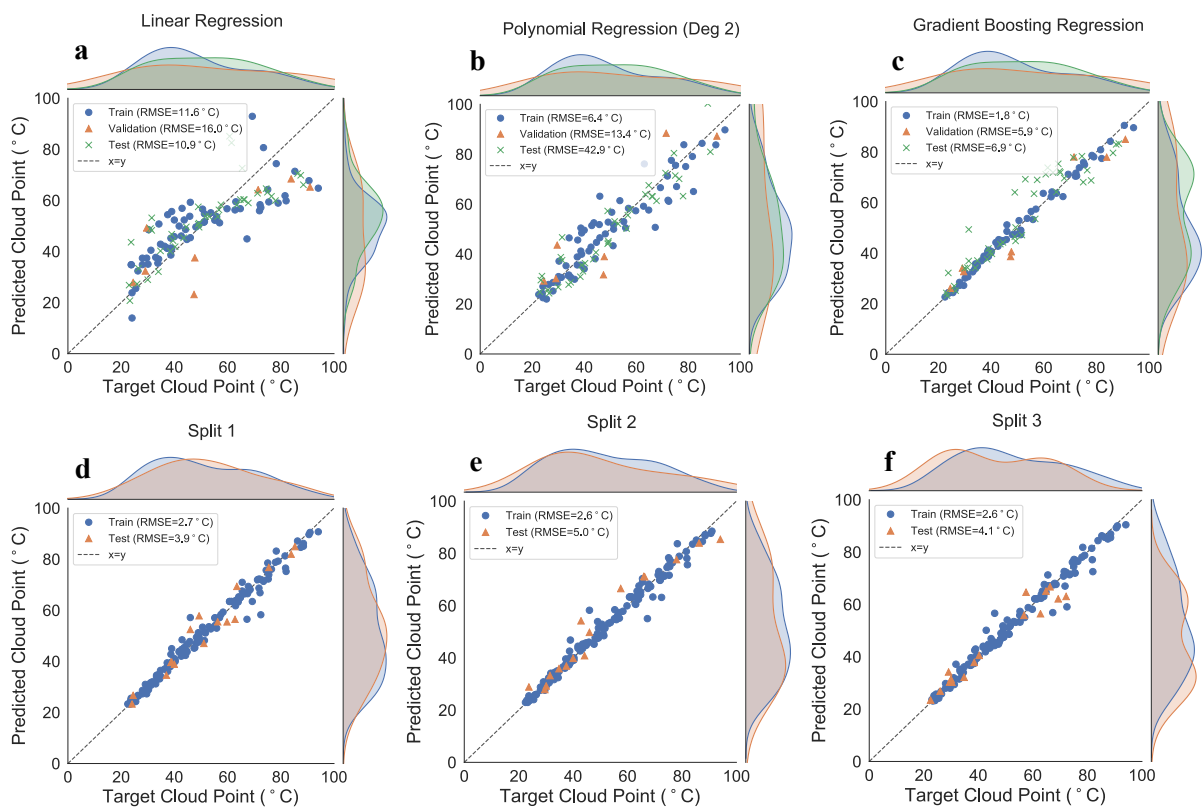

**Figure 3**: (top row) Comparison of three regression methods (a-c: linear, polynomial (order-2), and gradient boosting (with decision trees) regressions.) The literature data is split into 68 training data points and 7 validation data points. Test datapoints are 42 experimental data points produced in the lab. The results were compared using the root-mean-squared error. We observe that GBR achieves the best generalization. (bottom row) Final GBR model performance on 3 different random train-test splits of the combined dataset.

Linear and polynomial regressions, while significantly faster than the others, performed poorly when compared to SVR, NN and GBR. Of the latter three, GBR was the more accurate "out of the box". Moreover, it possesses fast inference speed, which is essential for efficient exploration of the parameter space in inverse design. We increased the predictive accuracy by tuning *via* a cross-validation grid search on hyper-parameters. We used both historic and experimental data, with a test set of 10% in order to validate our choice of hyper-parameters with the test error on 3 randomly split training and test sets (**Figure 3**). We now observe improved performance with an increased dataset and thorough tuning.

This algorithm is shown to generalize well across the variation in polymer dataset of varying polydispersity. The historical datasets had narrow polydispersity indices with the assumption of symmetrical MWDs, while the synthesized polymers had broad and unsymmetrical MWDs. The robustness of this algorithm in handling "noisy" data renders this far more powerful than a simple algorithm which only works for the highest quality of data. With a sufficiently accurate model, we finally retrain (using the tuned hyper-parameters) on the entire dataset to produce a finalized forward model that we use for subsequent inverse design.

While a forward predictive models in machine learning approaches for materials science are fairly common, inverse design is far more challenging. This is because the descriptors, which are usually high dimensional, are difficult to predict from outputs which are low dimensional. In the case of our polymer dataset, the output of cloud point is a single number, attributed to the 5 numbers representing molecular mass and composition of the polymer.

Inverse design would provide the ability to design polymers based on a desired final property and accelerate the synthesis process of target polymers based on design constraints to meet desired cloud points. To further realize new material discovery, we propose to extrapolate from our training dataset by designing terpolymers, which are non-existent in our training set, and limiting EtOx composition which is common.

Typically, inverse optimization on piece-wise constant functions provides a large number of different predicted designs. These may achieve our optimization and constraint target according to the fitted GBR model. However, the quality of these designs vary, particularly in the case of extrapolation. Validating all of these experimentally would be inefficient and so a filtering method with an ensemble of $M$ three-layer fully connected neural networks (NN) was employed to select the most promising design candidates for experimental validation. Each NN's trainable parameters are initialized with distinct, random values, resulting in different fitted predictors $\{\hat{f}_1, \ldots, \hat{f}_M\}$, due to the non-convex nature of the objective function and random initialization. For each design $x$, we then compared the ensemble of NN-predicted cloud points $\{\hat{f}_1(x), \ldots, \hat{f}_M(x)\}$ with the GBR prediction $\hat{f}(x)$ and only experimentally validated designs where $\hat{f}(x) \approx \frac{1}{M}\sum_{i=1}^{M} \hat{f}_i(x)$ (NN predictions agree with GBR) and $Var\{\hat{f}_1(x), \ldots, \hat{f}_M(x)\}$ was small. This ensures that $x$ is predicted with high confidence and not an ad-hoc extrapolation. **Figure 4** illustrates the principle of this approach. Although the NNs are also good approximators for the cloud point, they were not used as the forward model for producing inverse design candidates because the feed-forward step of the NN ensemble is still too slow compared with GBR, which consists of simple summing of piecewise constant functions.

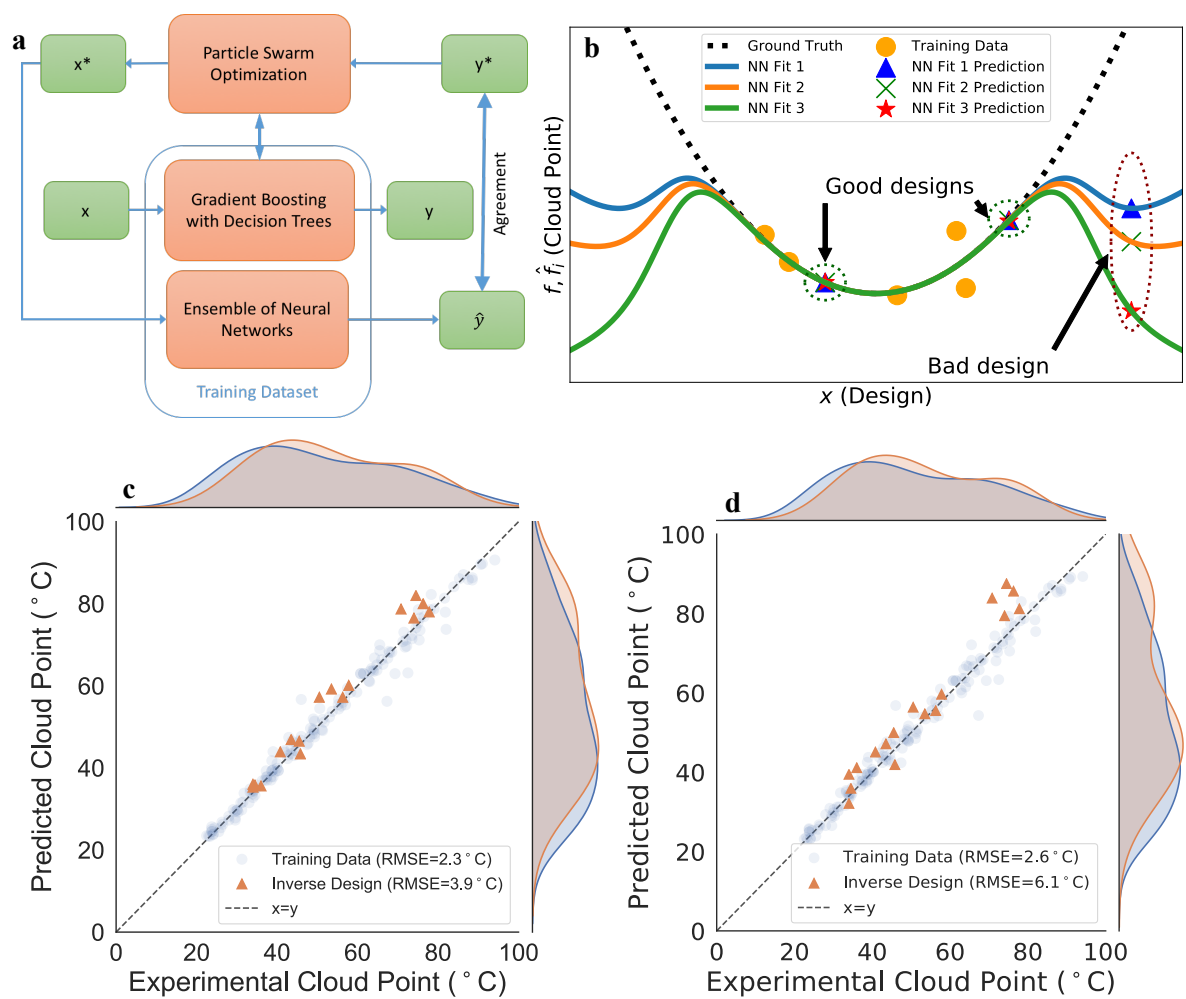

Figure 4: (a) Framework of the selection criteria, where the dataset is used to train GBR and NN ensemble, PSO predicts polymer design(x*) a desired CP(y*), and the design is verified for accuracy by the NN ensemble where CP agreement is a down-selection criteria (b) Illustration of the validity of the filtering procedure. We observe that given limited training data, not all extrapolated points are valid. However, when an ensemble of neural networks trained with distinct initializations agree on a certain input, then we have a much greater confidence in the validation of their predictions. (c) Final PSO based inverse design performance, showing an RMSE of 3.9 ˚C; (d) Forward model (NN ensemble) performance of the polymers synthesized from design

Using this technique, we down-selected 17 polymers over our 4 desired cloud-points (37, 45, 60, 80 ˚C), and imposed design criteria weighted on minimizing EtOx and designing polymers with more than two components – unseen in the training data. These polymers were synthesized, although an average of 3 iterations were required to achieve the target mass and composition of the designs, owing to the difficulties with terpolymer synthesis, where the Mayo-Lewis equation does not apply in calculating required feed ratio of monomer for desired final copolymer composition. The mass and composition of the synthesized polymers are reported in **Table S4**, showing minimal deviation from algorithmic design, along with their cloud points (an average of 3 measurements). The RMSE of the obtained cloud points was 3.9 ˚C, however when the polymer structure of the new polymers is fed back into the NN ensemble, a larger RMSE is observed (6.1 ˚C) (**Figure 4**). Deviation from the target cloud points was within test RMSE between 37-60 ˚C but above it at 80 ˚C, and can be attributed to sparseness of the data set at higher temperatures (**Figure 2F**) – an in-depth analysis is provided in ESI. These results show that our combination of slow and fast algorithms are able to design polymers with unique compositions with control over the desired physical property and structural design.

Overall, a significant conceptual advance in polymer inverse design has been achieved *via* judicious application of machine learning methods. This was done in three steps. First, we curated and categorized historical and new data. Second, we selected and fine-tuned a machine learning model based on gradient boosting regression with decision trees, resulting in a cloud point predictive accuracy of 3.9 ˚C (RMSE). The model was able to generalize well with both well-defined historic datasets as well as newly synthesized polymers of unsymmetrical MWDs. Third, inverse design by particle swarm optimization which predicted the design of new polymers based on desired cloud points (37, 45, 60, 80 ˚C). Extrapolation beyond the training set was achieved *via* an ensemble of neural-networks as a cross-validation technique to down-select 17 polymers with the lowest variance across predictions. The RMSE of predicted polymers were similar to those of the forward model. This methodology offers unprecedented control of polymer design, which may significantly accelerate the development of polymers with other physical properties.

Details of our code implementation and dataset can be found in our repository. (https://github.com/LiQianxiao/CloudPoint-MachineLearning)

All other information is in the ESI.

## Acknowledgements

J.K., Q.L., T.B. are support by the AME Programmatic Fund by the Agency for Science, Technology and Research under Grant No. A1898b0043

# Machine learning enables polymer cloud-point engineering *via* inverse design

Jatin N. Kumar, Qianxiao Li, Karen Y.T. Tang, Tonio Buonassisi, Anibal L. Gonzalez-Oyarce, Jun Ye

## Material synthesis and characterization

### Materials

2-*n*-propyl-2-oxazoline (nPropOx),(*1*) 2-cyclopropyl-2-oxazoline (cPropOx),(*2*) 2-isopropyl-2-oxazoline (iPropOx)(*3*) were synthesized as described in literature, and distilled over calcium hydride and stored with molecular sieves (size 5 Å) in a glovebox. 2-ethyl-2-oxazoline (EtOx, Sigma-Aldrich) was distilled over calcium hydride and stored with molecular sieves (size 5 Å) in glovebox. All other reagents were used as supplied unless otherwise stated.

### Analytical Methods

**Nuclear magnetic resonance (NMR).** The compositions of the polymers were determined using $^1$H NMR spectroscopy. $^1$H NMR spectra were on JEOL 500MHz NMR system (JMN-ECA500IIFT) in CDCl$_3$. The residual protonated solvent signals were used as reference.

**Size exclusion chromatography (SEC).** Gel permeation chromatography (GPC) measurements were performed in THF (flowrate: 1 mL/min) on a Viscotek GPC Max module equipped with Phenogel columns ($10^{-3}$ and $10^{-5}$ Å) (size: 300 x 7.80 mm) in series heated to 40 ºC. The average molecular weights and polydispersities were determined with a Viscotek TDA 305 detector calibrated with poly(methyl methacrylate) standards.

**Dynamic Light Scattering (DLS).** Measurements at various temperatures were conducted using a Malvern Instruments Zetasizer Nano ZS instrument equipped with a 4 mV He–Ne laser operating at l = 633 nm, an avalanche photodiode detector with high quantum efficiency, and an ALV/LSE-5003 multiple tau digital correlator electronics system. on Malvern Nano ZS. Solutions of polymers (5 mg/mL) were prepared by dissolving polymer in deionized water at room temperature. The solutions were then heated to 100 °C and cooled down to remove thermal memory, before measurements were taken.

### Experiments

For all polymerizations, the polymerization mixture was prepared in vials that were dried in 100 °C oven overnight before use, and crimped air-tight in a glove box. The mixture contained the monomers (EtOx, nPropOx, cPropOx, iPropOx) of desired ratios, with a total monomer concentration of 4 M, anhydrous acetonitrile (ACN) and methyl tosylate (MeOTs) as initiator. The amount of methyl tosylate added was determined by the various [M]/[I] ratios. Temperature controlled polymerizations were performed in sealed vials in a microwave reactor equipped with IR temperature sensor at 140 °C for different length of time. The mixture was then cooled to ambient temperature and quenched by addition of tetramethylammonium hydroxide (2.5wt% in methanol, 2 equivalence relative to initiator). The solutions were concentrated by removing some of the solvent under reduced pressure, then precipitated in cold diethyl ether. The product was collected and dried under reduced pressure overnight. All polymers were redissolved in THF for SEC, $CDCl_3$ for $^1H$ NMR and deionized water for DLS. $^1H$ NMR of $P((EtOx)_w(nPropOx)_x(cPropOx)_y(iPropOx)_z)$ (500 MHz, $CDCl_3$, $\delta$, ppm): 0.8 (*d*, 66.5 Hz, 4*y* H, CHC*H₂* C*H₂*), 0.96 (*s*, 3*x* H, $CH_2CH_2CH_3$), 1.11 (*s*, 6*z* H, $CHCH_3CH_3$), 1.12 (*s*, 3*w* H, $CH_2CH_3$), 1.64 (*s*, 2*x* H, $CH_2CH_2CH_3$) 2.30 (*d*, 56.5 Hz, 2*x* H, $NCOCH_2CH_2CH_3$), 2.38 (*s*, 2*w* H, $NCOCH_2CH_3$), 2.70 (*d*, 61.0 Hz, *y* H, C*H*CH₂CH₂), 2.80 (*d*, 123.5 Hz, *z* H, C*H*CH₃CH₃), 3.49 (*s*, 2(*w+x+y+z*) H, $CH_2$ backbone). Whereby *w*, *x*, *y* and *z* is the mole ratio of EtOx, nPropOx, cPropOx and iPropOx respectively.

# Curation and synthesis of the polymer library

To augment the historical dataset reported in **Table S1**, (*4-10*) a series of poly(2-oxazolines) were synthesized by cationic ring-opening polymerization in a microwave reactor at 140 °C and terminated with tetramethyl-ammonium hydroxide at the end of the reaction. All copolymers were synthesized with EtOx and one of the propyl oxazolines and variations in feed ratio were performed. SEC results are reported for all synthesized polymers in **Table S2**

**Table S1**: A list of historical data showing the degree of polymerization of EtOx (A), nPropOx (B), cPropOx (C), iPropOx (D), esterOx (E), polymer type (1 for homopolymer, 2 for statistical copolymer, 3 for gradient copolymer, 4 for block copolymer), molecular weight (M) in Da, polydispersity index (PDI), and cloud point in °C.

| No | Units of A | Units of B | Units of C | Units of D | Units of E | Type | M | PDI | Cloud Point |
|---|---|---|---|---|---|---|---|---|---|
| 1 | 10 | 0 | 0 | 0 | 0 | 1 | 1300 | 1.09 | - |
| 2 | 20 | 0 | 0 | 0 | 0 | 1 | 2000 | 1.08 | - |
| 3 | 30 | 0 | 0 | 0 | 0 | 1 | 2600 | 1.09 | - |
| 4 | 50 | 0 | 0 | 0 | 0 | 1 | 3800 | 1.09 | - |
| 5 | 100 | 0 | 0 | 0 | 0 | 1 | 6700 | 1.15 | 90.6 |
| 6 | 150 | 0 | 0 | 0 | 0 | 1 | 9000 | 1.15 | 85.3 |
| 7 | 200 | 0 | 0 | 0 | 0 | 1 | 13300 | 1.25 | 78.3 |
| 8 | 300 | 0 | 0 | 0 | 0 | 1 | 21000 | 1.33 | 73.5 |
| 9 | 500 | 0 | 0 | 0 | 0 | 1 | 37300 | 1.6 | 69.3 |
| 10 | 0 | 15 | 0 | 0 | 0 | 1 | 3100 | 1.1 | 42.9 |
| 11 | 0 | 20 | 0 | 0 | 0 | 1 | 3700 | 1.11 | 39 |
| 12 | 0 | 25 | 0 | 0 | 0 | 1 | 4300 | 1.14 | 37.5 |
| 13 | 0 | 50 | 0 | 0 | 0 | 1 | 6200 | 1.28 | 30.3 |
| 14 | 0 | 100 | 0 | 0 | 0 | 1 | 8140 | 1.4 | 29.6 |
| 15 | 0 | 150 | 0 | 0 | 0 | 1 | 12300 | 1.3 | 25.5 |
| 16 | 0 | 200 | 0 | 0 | 0 | 1 | 15500 | 1.43 | 24.1 |
| 17 | 0 | 300 | 0 | 0 | 0 | 1 | 18000 | 1.46 | 22.5 |
| 18 | 50 | 0 | 0 | 0 | 0 | 1 | 3300 | 1.14 | - |
| 19 | 45 | 5 | 0 | 0 | 0 | 2 | 3500 | 1.15 | - |
| 20 | 40 | 10 | 0 | 0 | 0 | 2 | 3500 | 1.36 | - |
| 21 | 35 | 15 | 0 | 0 | 0 | 2 | 3700 | 1.36 | 82 |
| 22 | 30 | 20 | 0 | 0 | 0 | 2 | 3700 | 1.34 | 72.2 |
| 23 | 25 | 25 | 0 | 0 | 0 | 2 | 4000 | 1.36 | 59.8 |
| 24 | 20 | 30 | 0 | 0 | 0 | 2 | 5400 | 1.35 | 51.3 |
| 25 | 15 | 35 | 0 | 0 | 0 | 2 | 3900 | 1.35 | 45.8 |
| 26 | 10 | 40 | 0 | 0 | 0 | 2 | 3800 | 1.34 | 40 |
| 27 | 5 | 45 | 0 | 0 | 0 | 2 | 4000 | 1.34 | 34.2 |
| 28 | 0 | 50 | 0 | 0 | 0 | 1 | 4200 | 1.32 | 29.6 |
| 29 | 100 | 0 | 0 | 0 | 0 | 1 | 15300 | 1.21 | 94.1 |
| 30 | 90 | 10 | 0 | 0 | 0 | 2 | 15200 | 1.22 | 81.6 |
| 31 | 80 | 20 | 0 | 0 | 0 | 2 | 13600 | 1.21 | 75.5 |
| 32 | 70 | 30 | 0 | 0 | 0 | 2 | 12600 | 1.26 | 64.8 |
| 33 | 60 | 40 | 0 | 0 | 0 | 2 | 13000 | 1.25 | 55.9 |
| 34 | 50 | 50 | 0 | 0 | 0 | 2 | 10700 | 1.28 | 51.1 |
| 35 | 40 | 60 | 0 | 0 | 0 | 2 | 10200 | 1.37 | 44.2 |
| 36 | 30 | 70 | 0 | 0 | 0 | 2 | 9700 | 1.36 | 40 |
| 37 | 20 | 80 | 0 | 0 | 0 | 2 | 9600 | 1.37 | 34.8 |
| 38 | 10 | 90 | 0 | 0 | 0 | 2 | 7800 | 1.48 | 31.2 |
| 39 | 0 | 100 | 0 | 0 | 0 | 1 | 8140 | 1.4 | 28 |
| 40 | 150 | 0 | 0 | 0 | 0 | 1 | 17700 | 1.47 | 83.9 |
| 41 | 135 | 15 | 0 | 0 | 0 | 2 | 17000 | 1.44 | 71.5 |
| 42 | 120 | 30 | 0 | 0 | 0 | 2 | 17700 | 1.4 | 63.1 |
| 43 | 105 | 45 | 0 | 0 | 0 | 2 | 17200 | 1.49 | 53.7 |
| 44 | 90 | 60 | 0 | 0 | 0 | 2 | 17900 | 1.38 | 49.2 |
| 45 | 75 | 75 | 0 | 0 | 0 | 2 | 17600 | 1.37 | 42.6 |
| 46 | 60 | 90 | 0 | 0 | 0 | 2 | 18600 | 1.35 | 37.3 |
| 47 | 45 | 105 | 0 | 0 | 0 | 2 | 17900 | 1.33 | 34.1 |
| 48 | 30 | 120 | 0 | 0 | 0 | 2 | 18500 | 1.34 | 29.1 |
| 49 | 15 | 135 | 0 | 0 | 0 | 2 | 18600 | 1.35 | 24.5 |
| 50 | 0 | 150 | 0 | 0 | 0 | 1 | 17200 | 1.45 | 24.1 |

| | | | | | | | | |
|---|---|---|---|---|---|---|---|---|
| 51 | 0 | 0 | 100 | 0 | 0 | 3 | 19700 | 1.18 | 28 |
| 52 | 11 | 0 | 89 | 0 | 0 | 3 | 20500 | 1.16 | 33 |
| 53 | 20 | 0 | 80 | 0 | 0 | 3 | 18000 | 1.16 | 39 |
| 54 | 31 | 0 | 69 | 0 | 0 | 3 | 14600 | 1.11 | 46 |
| 55 | 50 | 0 | 50 | 0 | 0 | 3 | 13400 | 1.14 | 57 |
| 56 | 69 | 0 | 31 | 0 | 0 | 3 | 15400 | 1.11 | 72 |
| 57 | 78 | 0 | 22 | 0 | 0 | 3 | 14100 | 1.1 | 79 |
| 58 | 100 | 0 | 0 | 0 | 0 | 1 | 14000 | 1.19 | 91 |
| 59 | 100 | 0 | 0 | 0 | 0 | 1 | 8000 | 1.02 | - |
| 60 | 22 | 0 | 0 | 78 | 0 | 3 | 9300 | 1.02 | 67.3 |
| 61 | 48 | 0 | 0 | 52 | 0 | 3 | 9300 | 1.02 | 55.2 |
| 62 | 73 | 0 | 0 | 27 | 0 | 3 | 9300 | 1.02 | 46 |
| 63 | 0 | 0 | 0 | 100 | 0 | 1 | 9700 | 1.02 | 38.7 |
| 64 | 0 | 0 | 0 | 40 | 10 | 4 | 6098 | | 44.7 |
| 65 | 0 | 0 | 0 | 40 | 20 | 4 | 7670 | | 47.7 |
| 66 | 0 | 0 | 0 | 40 | 40 | 4 | 10813 | | 47.4 |
| 67 | 0 | 106 | 0 | 0 | 0 | 1 | 12000 | 1.04 | 23.8 |
| 68 | 0 | 94 | 0 | 24 | 0 | 3 | 13300 | 1.03 | 26.3 |
| 69 | 0 | 59 | 0 | 50 | 0 | 3 | 12300 | 1.02 | 30.1 |
| 70 | 0 | 36 | 0 | 73 | 0 | 3 | 12300 | 1.02 | 33.8 |
| 71 | 0 | 0 | 0 | 86 | 0 | 1 | 9700 | 1.02 | 38.7 |
| 72 | 0 | 106 | 0 | 0 | 0 | 1 | 12000 | 1.04 | 23.8 |
| 73 | 34 | 84 | 0 | 0 | 0 | 3 | 12900 | 1.02 | 36.3 |
| 74 | 59 | 58 | 0 | 0 | 0 | 3 | 12400 | 1.02 | 41.8 |
| 75 | 96 | 40 | 0 | 0 | 0 | 3 | 14000 | 1.05 | 50.6 |
| 76 | 94 | 8 | 0 | 0 | 0 | 3 | 10200 | 1.04 | 75.1 |
| 77 | 81 | 0 | 0 | 0 | 0 | 1 | 8000 | 1.02 | - |
| 78 | 0 | 0 | 0 | 17 | 0 | 1 | 1900 | 1.05 | 72.5 |
| 79 | 0 | 0 | 0 | 21 | 0 | 1 | 2400 | 1.04 | 62.8 |
| 80 | 0 | 0 | 0 | 41 | 0 | 1 | 4600 | 1.03 | 51.3 |
| 81 | 0 | 0 | 0 | 50 | 0 | 1 | 5650 | 1.03 | 48.1 |
| 82 | 0 | 0 | 0 | 38 | 0 | 1 | 4300 | 1.03 | 43.7 |

| | | | | | | | | |
|---|---|---|---|---|---|---|---|---|
| 83 | 0 | 0 | 0 | 69 | 0 | 1 | 7800 | 1.02 | 38.7 |
| 84 | 0 | 0 | 0 | 86 | 0 | 1 | 9700 | 1.02 | 37.3 |

DLS measurements were performed in triplicate by preparing solutions of polymers at a concentration of 5 mg/mL in deionized water. The solutions were then heated to 100 °C and cooled down before measurements were taken to negate effect of thermal history. DLS measurements of the polymer solutions were performed over a temperature sweep between 20 to 90 ˚C. The cloud point temperature for the synthesized polymers (**Table S2**) was determined as the temperature at which the dissolved polymer chains of small hydrodynamic diameter agglomerate to form large particles or mesoglobules, as demonstrated in **Figure S1** for poly(nPropOx-co-EtOx) copolymers with a compositional variation at 20% increments.

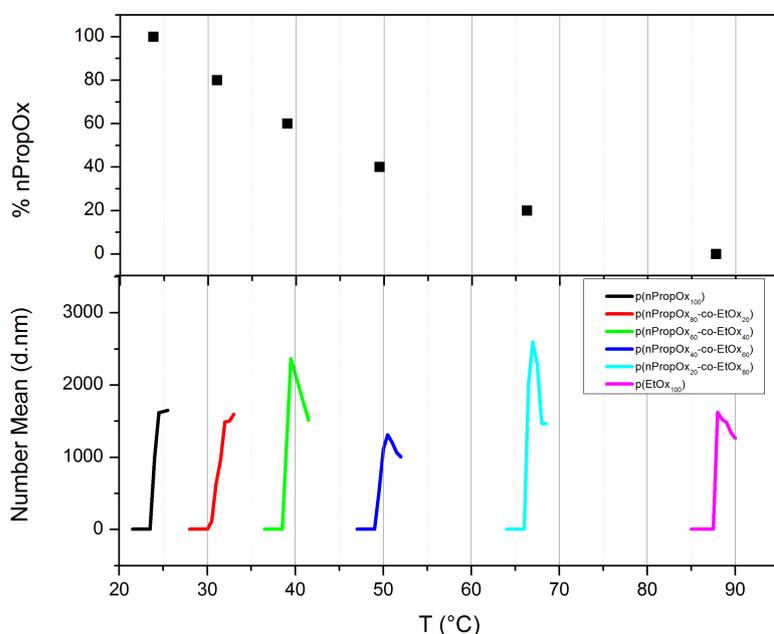

**Figure S1**: Temperature dependent DLS measurements for poly(nPropOx-co-EtOx) at various compositional ratios demonstrating the cloud point dependence on polymer composition.

**Table S2**: A list of data for synthesized polymers showing the degree of polymerization of EtOx (A), nPropOx (B), cPropOx (C), iPropOx (D), esterOx (E), polymer type (1 for homopolymer, 2 for statistical copolymer, 3 for gradient copolymer, 4 for block copolymer), molecular weight (M) in Da, polydispersity index (PDI), and cloud point in ˚C.

| No | Units of A | Units of B | Units of C | Units of D | Units of E | Type | M (Da) | PDI | CP |
|---|---|---|---|---|---|---|---|---|---|
| 1 | 47 | 0 | 0 | 0 | 0 | 1 | 4626 | 1.766 | - |

| 2 | 73 | 0 | 0 | 0 | 0 | 1 | 7274 | 2.072 | - |
| 3 | 26 | 0 | 0 | 0 | 0 | 1 | 2624 | 1.747 | - |
| 4 | 136 | 0 | 0 | 0 | 0 | 1 | 13458 | 2.94 | - |
| 5 | 13 | 0 | 0 | 0 | 0 | 1 | 1294 | 1.194 | - |
| 6 | 69 | 0 | 0 | 0 | 0 | 1 | 6853 | 2.546 | - |
| 7 | 208 | 0 | 0 | 0 | 0 | 1 | 20642 | 1.932 | 81.8 |
| 8 | 15 | 0 | 0 | 0 | 0 | 1 | 1448 | 1.228 | - |
| 9 | 23 | 0 | 0 | 0 | 0 | 1 | 2315 | 1.763 | - |
| 10 | 181 | 0 | 0 | 0 | 0 | 1 | 17897 | 2.341 | 87.5 |
| 11 | 91 | 0 | 0 | 0 | 0 | 1 | 9020 | 2.318 | 88.5 |
| 12 | 129 | 0 | 0 | 0 | 0 | 1 | 12787 | 2.294 | 88.5 |
| 13 | 171 | 0 | 0 | 0 | 0 | 1 | 16917 | 2.542 | 86.3 |
| 14 | 10 | 0 | 0 | 0 | 0 | 1 | 976 | 1.193 | - |
| 15 | 75 | 0 | 0 | 0 | 0 | 1 | 7407 | 2.673 | - |
| 16 | 435 | 0 | 0 | 0 | 0 | 1 | 43082 | 2.656 | 60.8 |
| 17 | 1166 | 0 | 0 | 0 | 0 | 1 | 115564 | 2.54 | 60.8 |
| 18 | 388 | 0 | 0 | 0 | 0 | 1 | 38506 | 2.426 | 61.5 |
| 19 | 954 | 0 | 0 | 0 | 0 | 1 | 94602 | 3.017 | 61.5 |
| 20 | 208 | 0 | 0 | 0 | 0 | 1 | 20654 | 2.283 | 65.5 |
| 21 | 222 | 0 | 0 | 0 | 0 | 1 | 21991 | 3.492 | 65.5 |
| 22 | 130 | 15 | 0 | 0 | 0 | 2 | 14534 | 2.011 | 74.5 |
| 23 | 129 | 20 | 0 | 0 | 0 | 2 | 15033 | 2.294 | 71.3 |
| 24 | 72 | 7 | 0 | 0 | 0 | 2 | 7953 | 2.074 | 75.0 |
| 25 | 55 | 15 | 0 | 0 | 0 | 2 | 7213 | 2.368 | 66 |
| 26 | 57 | 16 | 0 | 0 | 0 | 2 | 7513 | 1.813 | 66.3 |
| 27 | 116 | 29 | 0 | 0 | 0 | 2 | 14801 | 2.612 | 64.8 |
| 28 | 141 | 35 | 0 | 0 | 0 | 2 | 17920 | 1.719 | 63 |
| 29 | 90 | 40 | 0 | 0 | 0 | 2 | 13442 | 2.089 | 56.8 |
| 30 | 49 | 22 | 0 | 0 | 0 | 2 | 7430 | 2.191 | 57.5 |
| 31 | 76 | 57 | 0 | 0 | 0 | 2 | 14044 | 2.017 | 49.3 |
| 32 | 84 | 56 | 0 | 0 | 0 | 2 | 14681 | 1.933 | 49 |
| 33 | 100 | 69 | 0 | 0 | 0 | 2 | 17784 | 2.167 | 49.3 |
| 34 | 58 | 59 | 0 | 0 | 0 | 2 | 12421 | 2.197 | 44.5 |
| 35 | 85 | 82 | 0 | 0 | 0 | 2 | 17695 | 2.283 | 43.5 |
| 36 | 50 | 79 | 0 | 0 | 0 | 2 | 13956 | 2.19 | 39.5 |
| 37 | 43 | 67 | 0 | 0 | 0 | 2 | 11895 | 1.732 | 39 |
| 38 | 63 | 99 | 0 | 0 | 0 | 2 | 17431 | 2.232 | 38.5 |
| 39 | 63 | 99 | 0 | 0 | 0 | 2 | 17431 | 1.555 | 37.5 |
| 40 | 38 | 93 | 0 | 0 | 0 | 2 | 14232 | 2.496 | 34.8 |
| 41 | 50 | 128 | 0 | 0 | 0 | 2 | 19451 | 2.539 | 34 |
| 42 | 12 | 52 | 0 | 0 | 0 | 2 | 7020 | 2.388 | 31 |
| 43 | 13 | 55 | 0 | 0 | 0 | 2 | 7430 | 1.867 | 31.5 |
| 44 | 34 | 137 | 0 | 0 | 0 | 2 | 18942 | 2.324 | 30 |
| 45 | 37 | 146 | 0 | 0 | 0 | 2 | 20170 | 1.617 | 29.5 |
| 46 | 12 | 114 | 0 | 0 | 0 | 2 | 14074 | 1.825 | 26.5 |
| 47 | 14 | 140 | 0 | 0 | 0 | 2 | 17228 | 2.094 | 26 |
| 48 | 0 | 71 | 0 | 0 | 0 | 1 | 7981 | 2.193 | 23.8 |
| 49 | 0 | 76 | 0 | 0 | 0 | 1 | 8574 | 2.132 | 23.8 |
| 50 | 0 | 163 | 0 | 0 | 0 | 1 | 18419 | 2.506 | 23.3 |
| 51 | 0 | 138 | 0 | 0 | 0 | 1 | 15647 | 1.997 | 23.0 |
| 52 | 198 | 0 | 19 | 0 | 0 | 2 | 21729 | 2.044 | 73.5 |
| 53 | 68 | 0 | 14 | 0 | 0 | 2 | 8337 | 2.289 | 76.5 |
| 54 | 119 | 0 | 47 | 0 | 0 | 2 | 17039 | 2.027 | 64.2 |
| 55 | 49 | 0 | 24 | 0 | 0 | 2 | 7497 | 2.299 | 63.5 |
| 56 | 63 | 0 | 79 | 0 | 0 | 2 | 14997 | 2.118 | 49.3 |
| 57 | 24 | 0 | 37 | 0 | 0 | 2 | 6479 | 2.188 | 53 |
| 58 | 19 | 0 | 42 | 0 | 0 | 2 | 6510 | 1.994 | 42.8 |
| 59 | 4 | 0 | 17 | 0 | 0 | 2 | 2211 | 1.743 | 49 |
| 60 | 4 | 0 | 38 | 0 | 0 | 2 | 4553 | 1.818 | 37 |
| 61 | 0 | 0 | 37 | 0 | 0 | 1 | 4059 | 1.649 | 31.8 |
| 62 | 178 | 0 | 0 | 20 | 0 | 2 | 19822 | 1.921 | 75.5 |
| 63 | 70 | 0 | 0 | 11 | 0 | 2 | 8223 | 2.226 | 78 |
| 64 | 151 | 0 | 0 | 50 | 0 | 2 | 20623 | 1.926 | 63.8 |
| 65 | 49 | 0 | 0 | 21 | 0 | 2 | 7244 | 2.435 | 67.8 |

| 66 | 130 | 0 | 0 | 87 | 0 | 2 | 22756 | 1.978 | 55.3 |
| 67 | 40 | 0 | 0 | 44 | 0 | 2 | 8994 | 2.109 | 56.3 |
| 68 | 78 | 0 | 0 | 128 | 0 | 2 | 22154 | 1.93 | 48.3 |
| 69 | 14 | 0 | 0 | 35 | 0 | 2 | 5388 | 2.151 | 51.7 |
| 70 | 29 | 0 | 0 | 175 | 0 | 2 | 22614 | 1.863 | 43.5 |
| 71 | 0 | 0 | 0 | 83 | 0 | 1 | 9447 | 2.252 | 40.3 |
| 72 | 0 | 0 | 0 | 219 | 0 | 1 | 24748 | 1.823 | 38.5 |
| 73 | 108 | 199 | 0 | 0 | 0 | 2 | 33175 | 1.968 | 33.3 |
| 74 | 95 | 190 | 0 | 0 | 0 | 2 | 30964 | 1.685 | 33.3 |
| 75 | 29 | 0 | 0 | 188 | 0 | 2 | 24111 | 1.727 | 41.8 |
| 76 | 170 | 113 | 0 | 0 | 0 | 2 | 29644 | 1.924 | 44.8 |
| 77 | 109 | 130 | 0 | 0 | 0 | 2 | 25526 | 1.823 | 40.3 |
| 78 | 181 | 88 | 0 | 0 | 0 | 2 | 27894 | 1.871 | 48.7 |
| 79 | 142 | 0 | 66 | 0 | 0 | 2 | 21400 | 2.241 | 55.3 |
| 80 | 213 | 0 | 26 | 0 | 0 | 2 | 23961 | 2.111 | 68.5 |
| 81 | 291 | 10 | 0 | 0 | 0 | 2 | 29958 | 2.046 | 71.8 |
| 82 | 176 | 0 | 53 | 0 | 0 | 2 | 23380 | 2.238 | 64 |
| 83 | 224 | 0 | 31 | 0 | 0 | 2 | 25736 | 2.118 | 68 |
| 84 | 269 | 0 | 0 | 5 | 0 | 2 | 27290 | 1.971 | 75.3 |
| 85 | 1 | 0 | 38 | 0 | 0 | 2 | 4270 | 1.682 | 34.5 |
| 86 | 202 | 0 | 0 | 80 | 0 | 2 | 29092 | 2.104 | 58.3 |
| 87 | 251 | 0 | 0 | 23 | 0 | 2 | 27434 | 2.004 | 67.5 |

The PDIs obtained experimentally are much higher than the PDIs from the historical data. It can be assumed that the molecular weight distributions (MWD) for the historical data, where the PDI is lower than 1.4, are typically symmetrical. Conversely, the MWD of the polymers made experimentally had a long low-molecular weight tail (**Figure S2**). In the case of cationic ring opening polymerization, this long tail can be attributed to impurities such as water which terminate actively propagating chains. Due to the unsymmetrical MWD, the number average molecular weight ($M_n$), is no longer a proper representation of the MWD, particularly when comparing the dataset to historical data with polymers of narrow polydispersities.

Zhang et al.(11) propose that DLS is one of the better methods to characterize cloud points. They note that the intensity of scattered light due to a sharp change in refractive index is influenced by the chains that are dehydrating and thereby changing morphology from coil to globules. In contrast, only a minor difference in refractive index is observed from the hydrated chains. For broad or unsymmetric MWDs such as with our polymers, it seems intuitive that the cloud point by DLS of the modal polymer molecular weight would represent the polymer as a whole.

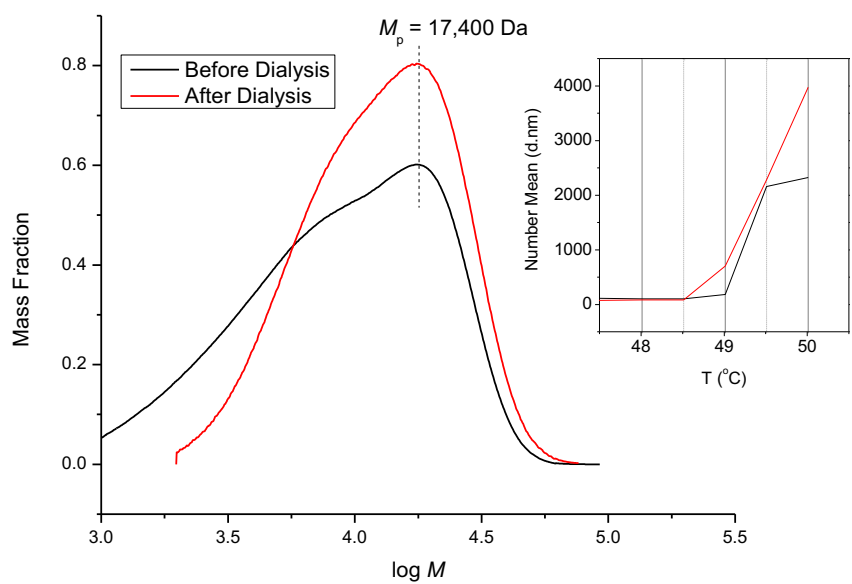

**Figure S2**: Gel permeation chromatogram and temperature dependent DLS data of poly(nPropOx-co-EtOx) (sample numbers 38 & 39, Table S2) before and after dialysis showing a narrowing of the molecular weight distribution, with no change in cloud point

To validate this theory, a polymer was selected at random, and dialyzed against water to remove some of the low molecular weight tail. Comparisons of the MWD before and after dialysis **(Figure S2)** show the removal of the low molecular weight tail, and the narrowing of the MWD. However, DLS results **(Figure S2-inset)** show no change to the cloud point of the polymer. Thus, to better represent the polymer dataset, the modal molecular weight, or peak molecular weight ($M_p$) was used to represent the molecular weight of the polymers from Table S2.

# Machine-Learning methodology

## Establishing a Machine Learning Baseline

It is often useful to establish a baseline for statistical methods on the currently available data before further data collection and algorithm exploration. In this section, we outline the development of our basic data driven approach which are broadly classified as statistical models (*e.g.*, multivariate analysis(*12*) and Bayesian inference(*13*)) and machine learning models (*e.g.*, support vector machines,(*14*) decision tree learning,(*15*) and deep neural networks(*16*)). The former perform well on relatively small datasets, but require non-trivial domain information such as statistical priors and a forward mathematical model, which may not always be available and can thus limit their applicability. On the other hand, machine learning models lend their applicability to datasets where the underlying physical mechanisms are unclear, or when the dataset has noise corruption.(*17*) While machine learning typically requires large datasets and cannot infer underlying physical relationships, its accuracy and fast inference speed makes it suitable for inverse design via global optimization.

In this work, we recall that we wish to predict the cloud point ($y \in R$) based on the polymer composition and other properties ($x \in R^d$). We assume that there is some relationship $y = f(x)$ for some unknown function $f$. Hence, our goal is to parameterize and fit an approximator $\hat{f}$ of $f$. The literature dataset is split into 68 training samples and 7 test samples, and we evaluate a total of five methods for fitting: 1) Linear regression; 2) Polynomial regression of degree up to 2; 3) Support vector regression; 4) Neural network regressor (2 hidden layers) 3) Gradient boosting regression with decision trees (GBR).(*17*) Below, we sketch the basic idea of the GBR method, which is the final choice of our forward model for inverse design and refer the reader to the text authored by Hastie, Tibshirani and Friedman(*18*) for more details.

## A Sketch of Gradient Boosting Regression

GBR makes use of the idea of "boosting", which is a class of sequential ensembling methods, where weak regressors (regression models with low capacity or approximation power) are iteratively combined to form a strong regressor. The basic idea is as follows: fix a space of weak approximators $H$ (e.g. decision trees) and start with a constant function $f_0$. For each $k \geq 0$, we set

$$f_{k+1} = f_k + \mathrm{argmin}_{h \in H} L(h, f - f_k) \qquad (1)$$

where the loss function $L$ measures the "distance" between its arguments. In other words, at each step we fit some function to approximate the current residual error $f - f_k$, and this successively improves the approximation. Of course, in practice the minimization step in (1) may be hard to evaluate, hence one can use "gradient boosting", where $h$ is not chosen as a true minimizer, but a function in the "steepest descent direction" of the loss function with respect to $h$. Detailed exposition on gradient boosting can be found int he previously mentioned text.(*18*)

The results of the comparisons are shown in **Figure 3** and **S3**, where we measure the root-mean-squared error on training, validation and test sets, the latter of which is the quantity to be used to discriminate model performances. The RMSE and the inference time is reported in **Table S3**. Note that while the training and validation sets are random splits in of the literature data, the test set are sample points obtained in our experiments. Thus, a model that performs well on the tests set indicates that it has the ability to fuse both literature data and our experimental data to form a more robust model. From our results, we observe that linear regression and polynomial regression, while having fast inference speeds, perform poorly in terms of test error. Moreover, polynomial regression suffers from the "curse of dimensionality" when higher order polynomials are included, since the number of terms increases exponentially with increasing maximum degree.

While all of the more sophisticated machine learning methods perform significantly better, the most outstanding is GBR method performs the best when weighing both in RMSE and inference time, even with minimal tuning. The inference time is important since we will need to repeated call this forward model in our inverse design process, and a faster inference time greatly enhances our exploration of the design space. Moreover, GBR (with decision trees as base regressors) give us a measure of feature importance using the Gini impurity.(*18*) In the present application, this gives us an estimation of the sensitivity of our cloud-point model on the polymer properties, seen in **Figure S3**.

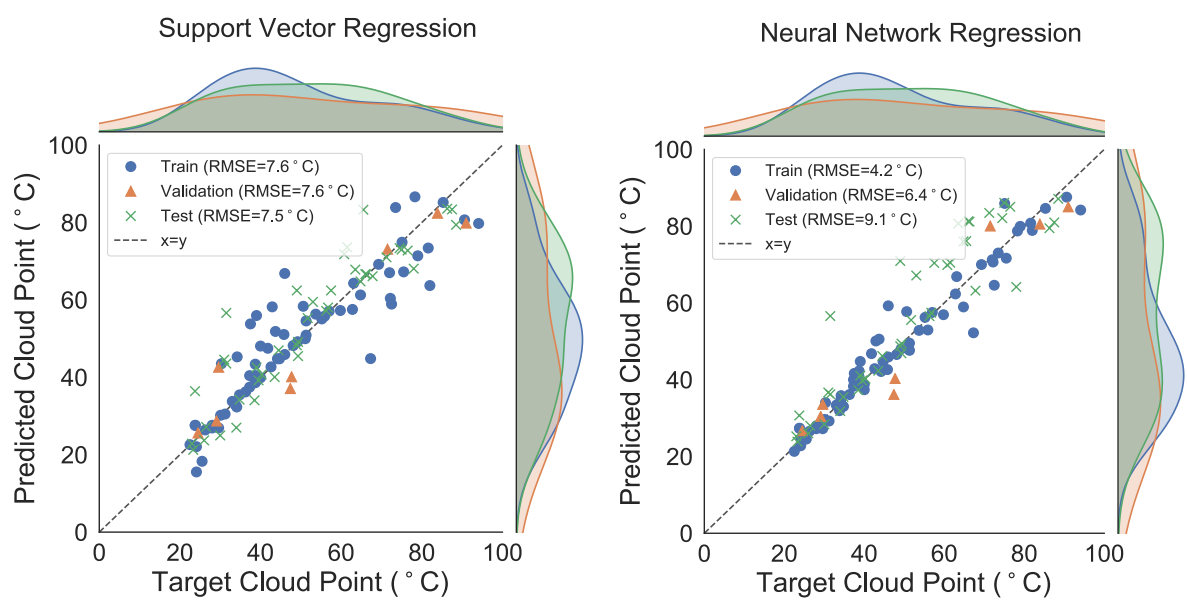

**Figure S3**: Comparison of two regression methods (support vector regression (SVR) and neural network regression (NN)). This serves as a basis of comparison to the other regressions shown in **Figure 3a-c**. The literature data is split into 68 training data points and 7 validation data points. Test datapoints are 42 experimental data points produced in the lab. The results were compared using the root-mean-squared error.

**Table S3**: A summary of the RMSE and inference times obtained by the 5 different regressions (linear, polynomial (degree 2), support vector, gradient boosting and 3 layer neural network)

| | Linear | Polynomial (°2) | Support Vector | Gradient Boosting | Neural Network |
|---|---|---|---|---|---|
| RMSE (°C) | 11.6±1.1 | 25.8±6.6 | 9.31±3.37 | 7.24±0.46 | 8.09±0.80 |
| Inference Time (μs) | 26.0±1.4 | 29.5±3.2 | 156 ± 13 | 161±9 | 235±14 |

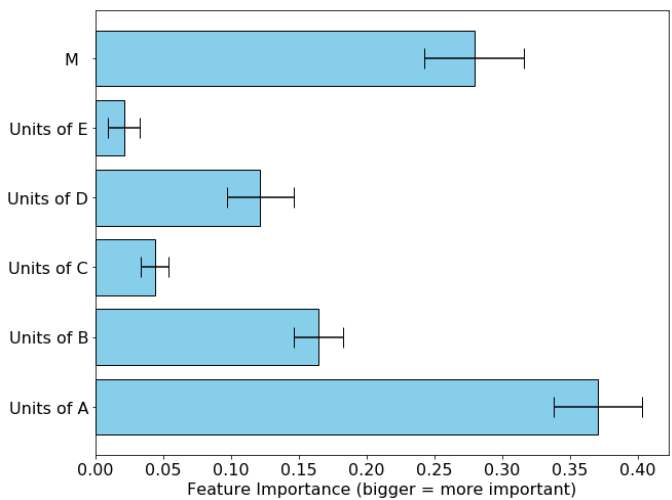

**Figure S4**: Feature importance via Gini impurity. Average values with standard deviation as error bars are plotted for each feature over 100 training-validation (90%-10%) splits

To optimize the GBR for inverse design, hyperparameter tuning was further conducted to bring the RMSE down to 3.9 °C. Details of which are presented in our data repository. With a tuned model, we look towards inverse design in order to predict polymer structure from desired cloud points.

Inverse Design via Particle Swarm Optimization

Our data-driven approximation $\hat{f}$ of the forward relationship between the polymer properties and the cloud point was demonstrated previously to be close to the true function $f$. In this section, we consider the problem of inverse design, where we want to find a polymer configuration $x$ that achieves certain targets (e.g. cloud point, desired proportions), while respecting certain constraints (e.g. molecular weight). Mathematically, this can be posed as a constrained optimization problem

$$\min_{x} F(x, \hat{f}(x)) \text{ subject to } G\left(x, \hat{f}(x)\right) \geq 0. \quad (2)$$

Where $F: R^d \times R \to R$ is the objective function and $G: R^d \times R \to R^m$ is the vector-valued constraint function.

The problem (2) is posed as a global optimization problem. In general, there are many heuristic methods for solving it, including simulated annealing,(*19*) genetic algorithms,(*20*) differential evolution,(*21*)etc. In this paper, we employ the particle swarm optimization (PSO) algorithm.(*22*) It is especially suited for our use-case since $\hat{f}$ is a boosted regression tree, which is a piece-wise constant function with almost everywhere vanishing derivatives, rendering gradient-based algorithms ineffective.

For the current application, we consider the following instance of objective and constraints:

Objective: Consider a mean-squared loss function that penalizes deviation from a target cloud point $y_t$ plus a regularization term that promotes certain desired design patterns

$$F\left(x, \hat{f}(x)\right) = \frac{1}{2}\left(\hat{f}(x) - y_t\right)^2 + R(x).$$

For the present application, we set $R$ so as to promote ternary and simpler designs (at most three non-zero components), as well as minimizing the units of A (EtOx). By writing $x = (x_A, x_B, x_C, x_D, x_E, x_M)$, we have

$$R(x) = \lambda_1 \sum_{\substack{x_1, x_2, x_3 \in (x_A, \dots, x_E) \\ x_1 \neq x_2 \neq x_3}} (x_1 x_2 x_3)^{1/3} + \lambda_2 x_A$$

where $\lambda_1, \lambda_2 > 0$ are regularization parameters. Note that there exist well-defined minima since we also require all components of $x$ to be non-negative.

Constraints: First, we employ the element-wise bounds
$$(0,0,0,0,0,0) \leq x \leq (203, 187, 43, 96, 0, 23196)$$
These bounds were selected based on the limits of the training data, and ease of synthesis.

Next, since M is the product of the degree of polymerization of the polymer and the molecular weight of its monomer units, we make sure that the designed $M$ values are consistent (within 10%) with the designed compositions, i.e.
$$0.8 * M(x) \leq x_M \leq 1.2 M(x)$$
where
$$M(x) = 99.13 x_A + 113.16 x_B, 111.14 x_C + 113.16 x_D$$

Finally, to simplify the experimental process we require the maximal number of monomer units to be at most 10 times of the minimal non-zero monomer unit, i.e.
$$\max_{i=A,\dots,E} x_i \leq 10 \min_{\substack{i=A,\dots,E \\ x_i > 0}} x_i$$

Our polymer design predictions were also given constraints based on our own requirements. For the purpose of this study, we chose to minimize the amount of EtOx in the polymer designs, especially since our training data was heavily populated with polymers containing EtOx. Thus, we ran four sets of predictions, in decreasing order of preference, where: (1) the algorithm limited EtOx to zero aggressively; (2) the algorithm limited EtOx to zero less aggressively; (3) the algorithm limited EtOx to under 100 units; (4) the algorithm did not limit EtOx. One of the other design parameters that we considered, was to have more than 2 components in the polymer design – a feature that was not present in our training set, nor is it commonplace when designing polymers for a desired physical property due to the expansion into a multivariable parameter space.

Selection Criteria

Besides obvious selection criteria such as picking designs with predicted cloud points close to the target cloud point, we developed more sophisticated selection procedures. As typical in inverse optimization on piece-wise constant functions, depending on the random initialization and the randomness of the PSO algorithm, we may arrive at a large number of different predicted designs that achieves, according to the fitted GBR model, our optimization and constraint targets. However, the quality of these designs varies (especially when extrapolating from our training data) and testing all of them would be inefficient. Thus, we employ a filtering method to select the most promising design candidates for experimental validation. Concretely, we train an ensemble of $M$ three-layer, fully connected neural networks (NN)(23) with sigmoid activations and mean-square loss on our full training set to predict cloud points based on polymer properties. Each NN's trainable parameters are initialized with distinct, random values. Due to the non-convex nature of the objective function and random initialization, with high probability each neural network will give rise to a different fitted predictor $\{\hat{f}_1, \ldots, \hat{f}_M\}$. For each design $x$, we then compare the ensemble of NN-predicted cloud points $\{\hat{f}_1(x), \ldots, \hat{f}_M(x)\}$ with the GBR prediction $\hat{f}(x)$. We only choose to experimentally validate designs where $\hat{f}(x) \approx \frac{1}{M}\sum_{i=1}^{M} \hat{f}_i(x)$ (NN predictions agree with GBR) and $Var\{\hat{f}_1(x), \ldots, \hat{f}_M(x)\}$ is small. This ensures that $x$ is predicted with high confidence and not an ad-hoc extrapolation. **Figure 4a and b** summarize and illustrate the principle of this approach. Note that although the NNs are also good approximators for the cloud point, we do not use NNs as the forward model for producing inverse design candidates because the feed-forward step of the NN ensemble is still too slow compared with GBR, which consists of simple summing of piecewise constant functions.

# Machine-Learning Validation

The inverse design generated a list of possible polymer mass and target compositions, following the 4 constraint parameters above, and are reported in their entirety in our code repository. The neural network on the trained dataset was used to predict a cloud point based on the particle swarm prediction of size and composition, and all predictions with the smallest difference between NN and GBR predictions and having low variance in the NN predictions were down-selected. From this, further down-selection was performed to choose 4 polymer designs per temperature with higher preference given to a more aggressive minimization of EtOx. The final choice of polymers is summarized in **Table S4**. It can be noted that almost all of the polymers were designed to have 3 components, with the exception of the 80 ˚C cloud point polymers.

Also in **Table S4** is the cloud point, composition and size of the polymers synthesized experimentally. The RMSE of the experimental results against their NN prediction was found to be 3.9 ˚C, which is the same the RMSE for the optimized GBR (**Figure 4C**). There is some deviation from the exact design due to experimental error, and when the obtained compositional and mass data was fed back into the NN for a forward predictive verification of the cloud points, a higher RMSE of 6.1 ˚C was seen (**Figure 4D**).

**Table S4**: A summary of the 17 polymers made, including their target cloud point and design along with the obtained cloud point and design

| Target CP (˚C) | Cloud point | | Mass | | | Target Composition | | | | Obtained Composition | | | |
|---|---|---|---|---|---|---|---|---|---|---|---|---|---|
| | Obtained CP (˚C) | Δ (˚C) | Target | Obtained | % Error | EtOx | nPropOx | cPropOx | iPropOx | EtOx | nPropOx | cPropOx | iPropOx |
| 37 | 34.5 | -2.5 | 13195 | 12629 | -4.3 | 28 | 79 | 0 | 13 | 28 | 78 | 0 | 9 |
| | 34 | -3 | 12191 | 12546 | 2.9 | 20 | 67 | 0 | 33 | 11 | 63 | 0 | 41 |
| | 36 | -1 | 10838 | 9629 | -11.2 | 21 | 59 | 0 | 40 | 30 | 52 | 0 | 33 |
| | 34 | -3 | 13875 | 14523 | 4.7 | 26 | 73 | 0 | 21 | 33 | 60 | 0 | 21 |
| 45 | 45.8 | 0.8 | 7712 | 6978 | -9.5 | 11 | 0 | 14 | 94 | 10 | 0 | 14 | 91 |
| | 43.5 | -1.5 | 10554 | 12151 | 15.1 | 26 | 0 | 22 | 72 | 21 | 0 | 15 | 79 |
| | 40.8 | -4.2 | 14496 | 14694 | 1.4 | 35 | 39 | 0 | 46 | 38 | 37 | 0 | 40 |
| | 45.5 | 0.5 | 7745 | 8040 | 3.8 | 26 | 0 | 28 | 65 | 24 | 0 | 18 | 72 |
| 60 | 57.8 | -2.2 | 20035 | 19901 | -0.7 | 84 | 0 | 16 | 20 | 80 | 0 | 12 | 23 |
| | 50.5 | -9.5 | 17111 | 17541 | 2.5 | 63 | 0 | 0 | 57 | 59 | 0 | 0 | 56 |
| | 53.5 | -6.5 | 11574 | 12447 | 7.5 | 73 | 38 | 9 | 0 | 70 | 39 | 6 | 0 |
| | 56.3 | -3.7 | 9257 | 8773 | -5.2 | 67 | 32 | 21 | 0 | 68 | 34 | 13 | 0 |
| 80 | 70.8 | -9.2 | 11725 | 10332 | -11.9 | 95 | 0 | 26 | 0 | 98 | 0 | 17 | 0 |
| | 74.5 | -5.5 | 18612 | 17021 | -8.5 | 104 | 0 | 17 | 0 | 103 | 0 | 12 | 0 |
| | 76.3 | -3.7 | 13330 | 13170 | -1.2 | 98 | 0 | 22 | 0 | 100 | 0 | 15 | 0 |

|   |   |   |   |   |   |   |   |   |   |   |   |   |
|---|---|---|---|---|---|---|---|---|---|---|---|---|
|   | 74 | -6 | 18975 | 17629 | -7.1 | 108 | 0 | 0 | 12 | 104 | 0 | 0 | 11 |
|   | 77.8 | -2.2 | 9079 | 9536 | 5.0 | 107 | 0 | 0 | 13 | 106 | 0 | 0 | 9 |

The higher RMSE can be attributed to a high variance for the polymers with a target cloud point of 80 °C. The RMSE from this method for just the polymers with a target cloud point of 37, 45 and 60 °C was 3.7 °C, while the RMSE for the polymers with a target cloud point of 80 °C was 9.1 °C (**Figure S5**). There are two reason for this: (1) fewer data points in this temperature range; (2) the algorithm minimizes A, which is typically required for higher cloudpoints.

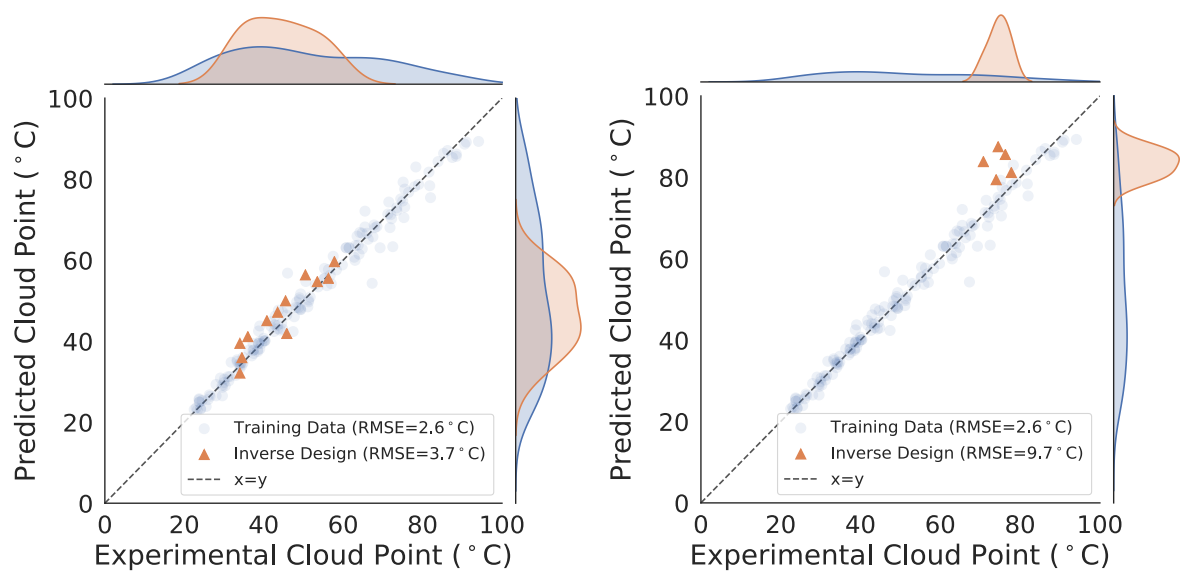

**Figure S5**: The fit of the NN ensemble model on experimentally obtained designs. The 80 °C designs are plotted separately from the other designs to show the main source of the deviation.

However, the results conclusively show that the inverse design algorithm is able to design polymers with unique compositions with a great deal of accuracy based on desired cloud points, especially when the cloud point range is well trained. The algorithm was robust enough to handle large variation in polymer quality as discussed earlier. Moreover, the algorithmic methodology allows us to vary our configuration for the inverse design, which would provide access a vast array of polymer design with a potential towards experiment automation. Lastly, the general nature of this algorithm could allow us to work with other similar polymer datasets, thereby accelerating the development of polymers in the future.